\documentclass[11pt,a4paper]{article}
\usepackage{jheppub}
\usepackage{etoolbox}
\usepackage{comment}

\def\one{{\,\hbox{1\kern-.8mm l}}}
\newcommand{\Dslash}{\not{\hbox{\kern-4pt $D$}}}
\newcommand{\pdslash}{\not{\hbox{\kern-2pt $\partial$}}}

\newcommand{\cM}{\mathcal{M}}

\newcommand{\Comment}[1]{{}}

\def\IZ{{\mathbb Z}}

\def\IR{{\mathbb R}}

\def\calh         {{\cal H}}

\def\calm         {{\cal M}}

\def\calo         {{\cal O}}

\newcommand{\bc}{\begin{center}}
\newcommand{\ec}{\end{center}}
\newcommand{\ba}{\begin{array}}
\newcommand{\ea}{\end{array}}
\newcommand{\beq}{\begin{equation}}
\newcommand{\eeq}{\end{equation}}
\newcommand{\bea}{\begin{eqnarray}}
\newcommand{\eea}{\end{eqnarray}}
\newcommand{\bmx}{\begin{pmatrix}}
\newcommand{\emx}{\end{pmatrix}}
\newcommand{\nn}{\nonumber}

\newcommand{\be}{\begin{equation}}
\newcommand{\ee}{\end{equation}}

\newcommand{\del}{\partial}
\newcommand{\half}{{\frac{1}{2}\,}}
\newcommand{\tr}{{\rm tr}}

\newcommand{\tD}{{\tilde D}}
\newcommand{\tdel}{{\tilde \del}}

\newcommand{\tm}{{\tilde m}}

\newcommand{\tpsi}{{\tilde \psi}}

\newcommand{\talpha}{{\tilde \alpha}}
\newcommand{\tN}{{\tilde N}}

\newcommand{\eref}[1]{Eq.\,(\ref{#1})}

\newcommand{\tchi}{{\tilde \chi}}
\newcommand{\tc}{{\tilde c}}
\newcommand{\tilh}{{\tilde h}}

\def\IB{\relax{\rm I\kern-.18em B}}
\def\IC{{\relax\hbox{\kern.3em{\cmss I}$\kern-.4em{\rm C}$}}}
\def\ID{\relax{\rm I\kern-.18em D}}
\def\IE{\relax{\rm I\kern-.18em E}}
\def\IF{\relax{\rm I\kern-.18em F}}
\def\II{\relax{\rm I\kern-.18em I}}
\def\IZ{\relax{\sf Z\kern-.35em Z}}
\def\Id{\relax{1\kern-.32em 1}}
\def\IG{\relax\hbox{$\inbar\kern-.3em{\rm G}$}}
\def\IR{\relax{\rm I\kern-.18em R}}
\newcommand\sfrac[2]{{\textstyle\frac{#1}{#2}}}

\title{Two-dimensional RCFT's without Kac-Moody symmetry}

\author{Harsha R. Hampapura$^a$}
\author{and Sunil Mukhi$^{a,b}$}

\affiliation{$^a$ Indian Institute of Science Education and Research,\\
Homi Bhabha Rd, Pashan, Pune 411 008, India}
\affiliation{$^b$ Yukawa Institute of Theoretical Physics\\
Kyoto University, Kyoto 606-8502, Japan}

\emailAdd{harshahr93@gmail.com}
\emailAdd{sunil.mukhi@gmail.com}

\abstract{Using the method of modular-invariant differential equations, we classify a family of Rational Conformal Field Theories with two and three characters having no Kac-Moody algebra. In addition to unitary and non-unitary minimal models, we find ``dual'' theories whose characters obey bilinear relations with those of the minimal models to give the Moonshine Module. In some ways this relation is analogous to cosets of meromorphic CFT's. The theory dual in this sense to the Ising model has central charge $\frac{47}{2}$ and is related to the Baby Monster Module.}

\preprint{YITP-16-61}

    
\begin{document}

\maketitle

\section{Introduction}

Conformal field theories in two dimensions \cite{DiFrancesco:1997nk} have a chiral symmetry algebra involving fields of spin $n\ge 1$. The conformal symmetry itself is generated by a spin-2 Virasoro algebra, while other algebras may or may not be present in general. In the holographic perspective \cite{Brown:1986nw} the spin-2 Virasoro algebra is the asymptotic symmetry of a spin-2 gauge theory, i.e. gravity, in the bulk. Likewise, a chiral algebra of spin-$N$ tells us that the bulk theory has spin-$N$ gauge fields. For $N=1$ this is the case of (Abelian or non-Abelian) gauge symmetry in the bulk. 

In this paper we will focus our attention on rational conformal field theories (RCFT's) in 2d, for which the total number of characters is finite. We will study the characters of these theories, defined as:
\be
\chi_i(q)=\tr_{{\cal H}_i} q^{-\frac{c}{24}+L_0}
\ee
where $q=e^{2\pi i\tau}$ and $\tau$ is the coordinate on the moduli space of the torus. The trace is taken over the Hilbert space $\calh_i$ of chiral states above the $i$th primary state. 
The characters are typically in correspondence with primary fields (upto degeneracies) of the full chiral algebra of the theory. Within RCFT's, the presence of spin-1 algebras is particularly interesting as they give rise to affine theories (WZW models, in Lagrangian language) based on a Kac-Moody current algebra. These in turn generate a vast set of 2d CFT's via the famous coset construction \cite{Goddard:1984vk, Goddard:1986ee}. 

Conversely one may ask for RCFT's that have no spin-1 chiral algebra. It is known \cite{Belavin:1984vu} that the only case where one can have a finite number of primaries without {\em any} other algebra beyond the Virasoro algebra, is when the central charge $c$ of this algebra is less than 1. The resulting models, called minimal models, are exactly solvable by virtue of having null vectors in the Virasoro module. They also have physical relevance, with the unitary series being related to RSOS models at criticality \cite{Andrews:1984af} and the $c=-\frac{22}{5}$  non-unitary model being related to the Lee-Yang edge singularity. The weaker condition that only a spin-1 algebra is absent (while arbitrary algebras of spin $N\ge 2$ are allowed) has not, to our knowledge, been classified. One interesting RCFT with this property is the ``Moonshine Module CFT'', which has a single character and is extremely interesting from the point of view of the mathematics of sporadic discrete groups \cite{Frenkel:1988xz}. Below we investigate possible solutions to the requirement of no Kac-Moody algebra for theories with a small number (but greater than 1) of characters. We will find new theories that bear an intriguing relationship with the Moonshine CFT.


One way to discover new RCFT's is to look for solutions to modular-invariant differential equations for their characters \cite{Anderson:1987ge,Eguchi:1988wh, Mathur:1988rx, Mathur:1988na, Mathur:1988gt}. In this approach one fixes a priori the number of characters of the desired theory, as well as an integer parameter $\ell$ describing the number of zeroes in moduli space of the Wronskian of the characters, and then writes down a general modular-invariant differential equation. For small $\ell$, this turns out to have a small number of arbitrary parameters. One then searches for those values of the parameters for which the solutions have a $q$-series expansion with non-negative integral coefficients. If this is verified to a sufficiently high order in the expansion then one has a candidate RCFT. With the available information, it is often possible to directly identify it as a WZW theory or coset theory \cite{Mathur:1988na, Hampapura:2015cea, Gaberdiel:2016zke}, thereby proving its existence as a CFT, and to reconstruct its fusion rules and primary correlators \cite{Mathur:1988gt}.

How is the absence of a spin-1 current algebra reflected in CFT characters? Among all the characters there is a unique one called the identity character. In unitary RCFT's this is the one with the most singular behaviour as $q\to 0$, though it can be more tricky to identify in non-unitary theories. Its leading behaviour is $q^{-\frac{c}{24}}$, where $c$ is the central charge of the theory. Let us now look at the coefficient of the successive term, $q^{-\frac{c}{24}+1}$. This is the number of states created from the ground state of the theory by acting on it with a mode of index $-1$ of one of the symmetry generators. If $K(z)$ is a generic spin-$N$ field in the chiral algebra then the modes are given by $K(z)=\sum_{n\in \IZ}K_n z^{-n-N}$. From considerations of non-singularity at the origin, it follows that:
\be
K_{-i}|0 \rangle=0,\qquad i<N
\ee
We see in particular that $K_{-1}|0\rangle=0$ for all generators of spin $N\ge 2$. Therefore the only way to have terms of order $q^{-\frac{c}{24}+1}$ in the identity character is to have currents $J^a_n$, and these terms must then have a $1-1$ correspondence with the states $J_{-1}^a|0\rangle$. Indeed, this approach was used in Refs.\cite{Mathur:1988na, Mathur:1988gt, Hampapura:2015cea, Gaberdiel:2016zke} to determine the dimension of the current algebra given the degeneracy of the first excited state in the identity character. It follows that if there are no spin-1 currents in the theory then the coefficient of $q^{-\frac{c}{24}+1}$ in the identity character must be zero. If we parametrise the identity character as:
\be
\chi_0(q)=q^{-\frac{c}{24}}\Big(1+m_1 q+m_2 q^2+\cdots\Big) 
\ee
then this tells us that $m_1=0$.

There is another way to have spin-1 currents, even if the identity character has $m_1=0$. Normally, characters are defined such that all states of integer dimension are included in the identity character. However it may happen that a theory has states of integer dimension but they are counted as descendants of some other primary (not the identity). Then they will appear in a distinct character built above a primary of integer dimension. If this primary has dimension 1 then it is a Kac-Moody current. Thus even with $m_1=0$ in the identity character, we have to ensure the absence of any dimension-1 primary in the theory by looking at the remaining characters.

To summarise, in order to discover potentially new RCFT's that do not have a Kac-Moody algebra, we have to look for consistent sets of characters transforming into each other under modular transformations and having the property that the identity character has $m_1=0$, and we also have to ensure there is no spin-1 primary in the theory.

Suppose first that the theory has a single character and a central charge of the form $c=24k$ where $k$ is an integer. Then of course the only requirement to have no Kac-Moody algebra is $m_1=0$. This can be explicitly solved as follows. The character must be a degree-$k$ polynomial in the Klein $j$-invariant. The $q$-expansion of $j$ is $j(q)=q^{-1}+744 + {\cal O}(q)$. We will find it more convenient to work with $J(q)=j(q)-744=q^{-1}+{\cal O}(q)$. Clearly a polynomial in $j$ is also a polynomial of the same degree in $J$. Now let us write:
\be
\chi_0(q)=P_k(J)=\sum_{m=0}^k a_m J^m 
\ee
In order for the identity field to be non-degenerate we must have $a_k=1$, and for the coefficient of $q^{-\frac{c}{24}+1}=q^{-k+1}$ to vanish we require $a_{k-1}=0$. These two conditions give us an infinite set of potential characters for one-character theories without Kac-Moody symmetry. The simplest theory in this class has $\chi(q)=J(q)$ and $c=24$, and corresponds to the famous Moonshine Module. In this example the CFT associated to this character is known to exist and has been constructed, but this is not yet the case for arbitrary characters of the above form. In general, we must think of the above conditions as necessary but not sufficient for the existence of CFT's without Kac-Moody symmetry.

Another interesting class of examples is provided by the $c<1$ minimal models. 
These are labelled by two integers $(p,p')$. Their central charge and primary conformal dimensions are as follows:
\be \label{minimalmod ch}
c = 1 - \frac{6(p-p')^2}{pp'} \hspace{1cm}  h_{r,s}=\frac{(rp'-sp)^2-(p-p')^2}{4pp'},\quad 1\le r \le p,~ 1\le s\le p'
\ee 
The unitary case corresponds to $p'=p+1$ and in this case the characters are given by \cite{DiFrancesco:1997nk}:
\be
\chi_{r,s}= K_{r,s} - K_{r,-s}
\ee 
where
\be
K_{r,s} = \frac{q^{-\frac{c}{24}} } {\prod_{n=1}^\infty (1-q^n)}  \sum_{n\in\IZ}  q^{\frac{(2np(p+1) +r(p+1) -sp)^2 -1}{4p(p+1)}}
\ee 
Using \eref{minimalmod ch}, the expression for the character becomes:
\be
\chi_{r,s} =\frac{q^{-\frac{c}{24} + h_{r,s} } } {\prod_{n=1}^\infty (1-q^n)} \sum_{n \in Z} q^{\frac{n}{2} } (1-q^{rs})
\ee
Evaluating the low-lying terms in the identity character, it is easy to verify explicitly that the first term above the ground state is absent. Indeed minimal models not only have no Kac-Moody symmetry, they have no other symmetry algebra besides the Virasoro algebra. For non-unitary minimal models the character formula above needs some modification, but the same conclusions hold.

We would now like to search for other RCFT's, besides those in the above examples, that have no Kac-Moody symmetries. For this we will use the method of modular-invariant differential equations, applied to RCFT's with small numbers of characters. We will re-discover the relevant minimal models, and also ``dual'' theories whose characters obey bilinear relations with those of the minimal models to give the character of the Moonshine CFT. We try to analyse what lessons can be learned from the existence of the latter types of characters. One of them, dual to the Ising model, has $c=\frac{47}{2}$ and has been previously studied by H\"ohn\cite{Hoehn:thesis}.

\section{Modular-invariant differential equations}

The characters of an RCFT arise as the independent solutions of a degree-$p$ modular-invariant differential equation in $\tau$. Such an equation must be of the form \cite{Anderson:1987ge,Eguchi:1988wh,Mathur:1988rx, Mathur:1988na, Mathur:1988gt}:
\be
\left(D^p + \sum_{k=0}^{p-1} \phi_k(\tau) D^k\right)\chi=0
\label{modinveq}
\ee
where $D$ is a covariant derivative to be defined below, and $\phi_k(\tau)$ is a modular function of weight $2(p-k)$ under SL(2,Z). The characters transform into each other under SL(2,Z) but they have zero weight. The derivative $D$ acting on them successively increases the weight by 2. It follows that every term in the above equation has modular weight $2p$, and the equation is therefore modular invariant. The covariant derivative is given by:
\be
D \equiv  \frac{\del}{\del\tau}-\frac{i\pi r}{6}E_2(\tau)
\label{covdev}
\ee
where $r$ is the modular weight of the object on which it acts, and $E_2(\tau)$ is a special Eisenstein series that transforms inhomogeneously under SL(2,Z) and thereby provides a suitable connection. 

In general $\phi_k$ need not be holomorphic, indeed  they can be meromorphic even though the resulting characters are holomorphic. In fact the poles of $\phi_k$ are related to the zeroes of the Wronskian of the independent solutions $\chi_0,\chi_1,\cdots \chi_{p-1}$ of the differential equation by the following relation:
\be
\phi_k(\tau)=(-1)^{n-k}\frac{W_k}{W}
\ee
where the Wronskian determinants $W_k$ are defined in Refs.\cite{Mathur:1988rx, Mathur:1988na, Mathur:1988gt}. In searching for new RCFT's one therefore starts by choosing the number of characters $p$ as well as the number of zeroes of $W$, which is of the form $\frac{\ell}{6}$ where $\ell$ is a non-negative integer other than 1 (fractional zeroes are allowed due to the orbifold singularities of the torus moduli space). The central charge and conformal dimensions of any RCFT satisfy the relation \cite{Mathur:1988na}:
\be
\sum_{i=0}^{p-1}\left(-\frac{c}{24}+h_i\right)=\frac{p(p-1)}{12}-\frac{\ell}{6}
\label{elldef}
\ee
It is tedious but straightforward to verify that all the $c<1$ minimal models, unitary or otherwise, have $\ell=0$. The same is true of SO(N) and SU(N) WZW models, but there are also many known CFT's with $\ell\ge 2$, some of which appear in known discrete series and others are constructed in Refs.\cite{Naculich:1988xv,Hampapura:2015cea,Gaberdiel:2016zke}. We now use this approach of modular-invariant differential equations to investigate the existence of CFT characters without Kac-Moody symmetries for small values of $\ell$.

\section{Theories without Kac-Moody symmetries}

\subsection{Two-character theories}

Let us start by fixing the number $\ell$ of zeroes of the Wronskian to be 0. The modular-invariant differential equation is simplest in this case, because the coefficient functions $\phi_k(\tau)$ are holomorphic everywhere in the interior of moduli space and therefore must be polynomials in the two Eisenstein series $E_4$ and $E_6$ (for definitions, see the Appendix).

The most general homogeneous, modular invariant, second order differential equation is:
\be \label{2char de}
\Big(\tD^2 + \phi_1(\tau)\tD+\phi_0(\tau)\Big)\chi=0
\ee
where $\tD=\frac{1}{2\pi i}D$ is the covariant derivative scaled for future convenience. 
Here $\phi_k$ are holomorphic and $\phi_1,\phi_0$ have modular weight $2,4$ respectively. It follows that $\phi_1=0$ and $\phi_0$ is proportional to $E_4$. Thus we have the differential equation:
\be
(\tD^2 + \mu E_4)\chi=0
\label{leq.0}
\ee
with $\mu$ a free parameter. In terms of ordinary derivatives this differential equation can be written as:
\be
\Big(\tdel^2-\frac16 E_2 \tdel + \mu E_4 \Big) \chi =0
\ee  
where $\tdel=\frac{1}{2\pi i}\del$.

In Ref. \cite{Mathur:1988na}, this equation was solved by substituting the mode expansions of the characters, $\chi=\sum_{n=0}^\infty a_n q^{\alpha+n}$, and the Eisenstein series $E_a(\tau)=\sum_{k=0}^\infty E_{a,k}\,q^k$. The result is 
the following set of equations. First of all, if $\alpha$ is either of the two exponents then
\be
\alpha^2-\sfrac16\alpha+\mu=0
\ee
Next, denoting the two roots of this equation by $\alpha_0,\alpha_1$ (where  $\alpha_0$ is the exponent corresponding to the identity character and $\alpha_1$ corresponds to the non-trivial primary), we have:
\be
\begin{split}
	\alpha_0+\alpha_1&=\frac16\\
	\mu = \alpha_0\alpha_1 &=\alpha_0\left(\frac16-\alpha_0\right)
\end{split}
\label{muval}
\ee
Note that $\alpha_0=-\frac{c}{24}$ and therefore our parametrisation of the identity character is:
\be
\chi_0=q^{\alpha_0}(1+m_1 q+ m_2 q^2+\cdots)
\ee
Going to next order in the series solution, and using the above results, we eventually get \cite{Mathur:1988na}:
\be
m_1=\frac{24\alpha_0(60\alpha_0-11)}{5+12\alpha_0}
\label{m.1.ell.0}
\ee

To find theories without Kac-Moody symmetries we set $m_1=0$ in the above equation and solve for $\alpha_0$. This directly gives $\alpha_0=\frac{11}{60}$. Identifying this with $-\frac{c}{24}$ gives us $c=-\frac{22}{5}$. This is a well-known minimal model corresponding to $(p,p')=(2,5)$. There are no more solutions to $m_1=0$ at $\ell=0$, beyond the trivial case $\alpha_0=0$ corresponding to $c=0$.

Next we look at the $\ell=2$ two-character theories, exhaustively studied in Ref.\cite{Hampapura:2015cea} (for earlier work on these theories see Ref.\cite{Naculich:1988xv}). In this case we will find something interesting. The value of $m_1$ for these theories is given by Eq.(5.26) of that reference in terms of another integer $\tN$. Putting $m_1=0$ one finds $\tN=144$, and indeed this value appeared in the list Eq.(5.27) of Ref.\cite{Hampapura:2015cea}. However this was then ruled out in that paper because the degeneracy at the second level above the identity turned out to be a {\em negative} integer. Computing the degeneracies of the identity character to very high powers in $q$, we find that except for the ground state, they are all negative integers. This makes it difficult to propose a physical meaning for this theory. However, it is still remarkable (and not the result of any prediction, since we do not know a candidate CFT for this case) that the degeneracies for the identity character are integral to very high orders in $q$ and we expect this property persists to all orders. The central charge of this would-be theory is found by setting $m_1=0$ in Eq(5.22) of Ref.\cite{Hampapura:2015cea}, from which one finds $c=\frac{142}{5}$.  Using the fact that $\ell=2$ one easily finds that the nontrivial primary of this theory has conformal dimension $\frac95$. Next the characters for this primary can be computed, upto normalisation, to any desired order in $q$ following the method of Ref.\cite{Hampapura:2015cea} and one finds that to very high orders the degeneracies are positive rational numbers with a denominator that appears to be bounded. Thus for this character a suitable degeneracy factor for the ground state would render it consistent. 

These empirical facts lead to an intriguing observation. The above ``characters'' bear a close relation to those of the non-unitary $c=-\frac{22}{5}$ minimal model via a bilinear relation, analogous to the one recently found in Ref.\cite{Gaberdiel:2016zke}. Let us exhibit the precise relationship. Denote the familiar $c=-\frac{22}{5}$ minimal model by ${\cal M}_{2,5}$ and let $\chi_0,\chi_1$ be its characters. Likewise, denote the (tentative) $\ell=2$ theory with $\tc=\frac{142}{5}$ as ${\tilde {\cal M}}_{2,5}$ and let $\tchi_0,\tchi_1$ be its characters. It is well-known that $\cM_{2,5}$ has a primary with $h=\frac15$, while we have just seen that ${\tilde \cM}_{2,5}$ has a primary of dimension $\tilh=\frac{9}{5}$. Putting all this together we find that $c+\tc=24$ and $h+\tilh=2$. This is precisely the relation between a specific affine theory and the coset of a meromorphic $c=24$ CFT by that affine theory, proposed in Ref.\cite{Gaberdiel:2016zke} and justified with numerous examples. However, there is an important difference. In the case of Ref.\cite{Gaberdiel:2016zke}, one really had a coset construction. The numerator theories were meromorphic $c=24$ CFT's having a Kac-Moody symmetry (not affine theories, but rather modular-invariant combinations of characters of a subset of the integrable primaries). The denominators were affine theories having a Kac-Moody algebra that is a direct summand of the one in the numerator. The Kac-Moody algebras play a crucial role in enabling a definition of these generalised cosets, as explained in detail in Ref.\cite{Gaberdiel:2016zke}. But in the present case there are no currents and therefore no coset construction. 

However, in analogy with Eq.(2.7) of Ref.\cite{Gaberdiel:2016zke} we can still look for a bilinear relation between the characters $\chi_i,\tchi_i$ described above and some single-character CFT. What theory should appear on the RHS of such a bilinear relation? It has to be a meromorphic $c=24$ CFT and must therefore appear in the list of Ref.\cite{Schellekens:1992db}. Since the pair of theories on the LHS have no Kac-Moody symmetry, the same property must hold for the RHS. There is a unique meromorphic CFT with this property, namely the famous Moonshine CFT whose character is $J(q)=j(q)-744$. Thus we are motivated to suggest a bilinear holomorphic relation as follows:
\be
\sum_{i=0}^{p-1}\chi_i(q)\tchi_i(q)=
j(q)-744
\label{jbilinear}
\ee
with $p=2$, the characters on the LHS being those of ${\cal M}_{2,5}$ and ${\tilde {\cal M}_{2,5}}$.
This is a precise, testable formula and can only hold if all the (infinitely many) coefficients in the $q$-series match on both sides. We know $\chi_i$, the characters of ${\cal M}_{2,5}$, and we can also use the modular differential equation to compute the characters $\tchi_i$ of the hypothetical theory ${\tilde {\cal M}_{2,5}}$ as explained above. Therefore we can check whether $\chi_i(q),\tchi_i(q)$ obey \eref{jbilinear} to any desired order in $q$. 

At leading order the relationship holds due to the matching of exponents discussed above. Once we go beyond that, there is a subtlety: so far, we do not know the degeneracy of the nontrivial primary whose character is $\tchi_1$. 
Let us assume all the characters under discussion are normalised so that the first term in unity. Let the degeneracy of the ground state be labelled by $D_0,D_1$ for the characters of $\cM_{2,5}$ and $\tD_0,\tD_1$ for those of ${\tilde\cM}_{2,5}$. Also let us use $\psi_i$ and $\tpsi_i$ to denote characters normalised so that the first term in the expansion is unity. We then have:
\be
\chi_i(q)=D_i\,\psi_i(q),\qquad \tchi_i=\tD_i\,\tpsi_i(q),\qquad i=0,1
\label{bilintwo}
\ee
One always has $D_0=\tD_0=1$ from non-degeneracy of the identity. Therefore in a general situation, the bilinear of interest is:
\be
\sum_{i=0}^{1}\chi_i(q)\tchi_i(q)=\psi_0(q)\tpsi_0(q)+D_1\tD_1\,\psi_1(q)\tpsi_1(q)
\ee
The expansion of $\psi_i$ is:
\be
\psi_i=q^{\alpha_i}\left(1+m_1^{(i)}q+m_2^{(i)}q^2+\cdots\right)
\ee
where $\alpha_0=-\frac{c}{24}$ and $\alpha_1=-\frac{c}{24}+h$. Note that the quantities previously called $m_1,m_2$ are now labelled $m_1^{(0)},m_2^{(0)}$ but we will revert to the simpler notation whenever there is no scope for confusion. A similar expansion holds for $\tpsi$.

Now from the above relations between the central charges and conformal dimensions of the paired theories, we have:
\be
\alpha_0+\talpha_0=-1, \qquad \alpha_1+\talpha_1=1
\ee
It follows that upto $\calo(q)$, \eref{bilintwo} is equal to:
\be
\begin{split}
&q^{-1}\left(1+m_1^{(0)}q+m_2^{(0)}q^2\right)\left(1+\tm_1^{(0)}q+\tm_2^{(0)}q^2\right)
+D_1\tD_1 q \\
&\quad = q^{-1}+(m_1+\tm_1)+(m_1\tm_1+m_2+\tm_2+D_1\tD_1)q
\end{split}
\ee
As promised, in the last line we have dropped the superscripts on $m_i,\tm_i$ because only those corresponding to the identity appear to this order. 

Now in the present case, $m_1=\tm_1=0$ and also $D_1=1$ because minimal models have non-degenerate primaries. 
Recall that:
\be
J(q)=j(q)-744=q^{-1} + 196884 q+\cdots
\ee
and therefore to satisfy \eref{jbilinear} we must have $\tD_1+m_2+\tm_2=196884$. Since $m_2$ and $\tm_2$ are directly calculable using the differential equation, this determines $\tD_1$. Applying it to the case of $\calm_{2,5}$ and $\tilde\calm_{2,5}$, we find that $m_2=1,\tm_2=-164081$. This determines $\tD_1=360964$. A non-trivial check of this normalisation is that in the expansion of the non-identity character of $\tilde\calm_{2,5}$, one finds fractional coefficients with denominators as large as 90241 (working up to $\calo(q^{1000})$). Thus it must be the case that 360964 is divisible by 90241, and this is true (the ratio is 4). This means that with this choice of $\tD_1$, the non-identity character of $\tilde\calm_{2,5}$ indeed has integer degeneracies.

Despite the above check, we have not yet performed any actual test of \eref{jbilinear}. But now all quantities on the LHS are known, as we can compute the power series for $\chi_i,\tchi_i$ to any desired order in $q$ and we have determined all the normalisations. We can then test \eref{jbilinear} order-by-order and we find that all the way to $\calo(q^{1000})$ it works perfectly.

To summarise, we have conjectured a bilinear relation between two  pairs of characters (one corresponding to a known non-unitary CFT and the other to a more exotic system with negative but integer degeneracies) to the Moonshine CFT, and verified this conjecture to $\calo(q^{1000})$. The significance of this construction is that it points the way to similar relations for {\em unitary} theories, where no negative degeneracies are present. Such relations cannot be sought within two-character theories because we do not know of any two-character {\em unitary} RCFT without a Kac-Moody algebra. However such theories do exist with $p\ge 3$ characters, where an infinite family is provided by the unitary minimal models, starting with the well-known Ising model. Hence we now turn our attention to the case $p=3$. We will repeat the procedures described above and find very analogous results. 

Note that, independent of the bilinear relation, we have successfully classified all possible two-character RCFT's with $\ell=0,2,3$ having no Kac-Moody algebra. For $\ell=0$ this is the Yang-Lee theory, for $\ell=2$ this is the exotic dual discussed above and for $\ell=3$ there are no candidates as shown in Ref.\cite{Hampapura:2015cea}.

\subsection{Three-character theories}

The case of three-character theories, even with $\ell=0$, is not completely classified despite non-trivial progress in Ref.\cite{Mathur:1988gt,Gaberdiel:2016zke}. It is known that infinitely many such theories exist, in sharp contrast to the case of two-character theories with $\ell=0$. However the best-known infinite series corresponds to the SO(N)$_k$ WZW models, which are not of interest to us here. We shall now re-open the investigation into three-character theories with $\ell=0$, but focusing specifically on solutions without Kac-Moody symmetry.

The modular invariant differential equation in this case is \cite{Mathur:1988gt}:
\be
\left(D_\tau^3 + \pi^2 \mu_1 E_4 D_\tau + i\pi^3 \mu_2 E_6\right)\chi(\tau)=0
\ee 
In terms of ordinary derivatives the above equation becomes:
\be \label{3char-de}
\Big(\partial^3_{\tau}-\frac{i\pi}{3}(\partial_{\tau}E_2)\partial_{\tau} -i\pi E_2\partial^2_{\tau} -\frac{2\pi^2}{9}E^2_2\partial_{\tau}+\mu_1 \pi^2 E_4\partial_{\tau}+i\mu_2 \pi^3 E_6\Big)\chi=0
\ee
As explained in Ref.\cite{Hampapura:2015cea}, one can use Ramanujan identities to make differential equations linear in the Eisenstein series. Accordingly, we apply the following identity to \eref{3char-de}:
\begin{equation}
\frac{1}{2i\pi}(\partial_{\tau}E_2) = \frac{E^2_2-E_4}{12}
\end{equation}
as a result of which the equation becomes:
\begin{equation}
\Big(\partial^3_{\tau}+i\pi(\partial_{\tau}E_2)\partial_{\tau} -i\pi E_2\partial^2_{\tau} -\frac{2\pi^2}{9}E_4\partial_{\tau}+\mu_1 \pi^2 E_4\partial_{\tau}+i\mu_2 \pi^3 E_6\Big)\chi=0
\end{equation}
Upon substituting the mode expansions we get the recursion relation:
\be
\begin{split} \label{recursion-3char}
	&-8(n+\alpha)^3 a_n- 4\sum_{k=0}^{n}E_{2,k} a_{n-k}
	k(n-k+\alpha)+ 4\sum_{k=0}^{n}(n-k+\alpha)^2a_{n-k}E_{2,k} \\ 
	&-\frac{4}{9} \sum_{k=0}^{n}  E_{4,k}(n-k+\alpha)a_{n-k} +2\mu_1 \sum_{k=0}^{n}(n-k+\alpha)E_{4,k}a_{n-k}+\mu_2\sum_{k=0}^{n}E_{6,k}a_{n-k}=0
\end{split}
\ee

For $n=0$ and $n=1$, we get the following polynomial equations in $\alpha$:
\be \label{3char-indicial}
-8\alpha^3 + 4\alpha^2 + -\frac{4}{9}\alpha + 2\mu_1\alpha+\mu_2=0
\ee
and
\be \label{3char-n1}
\begin{split}
&\left(-4E_{2,1}\alpha+4\alpha^2E_{2,1}-\frac{4}{9}E_{4,1}\alpha+\mu_2E_{6,1}+2\mu_1 E_{4,1}\alpha\right)\\
&\qquad\qquad\qquad + m_1\left(-24\alpha^2-16\alpha-\frac{40}{9}+2\mu_1\right)
=0
\end{split}
\ee

From these equations we immediately see that:
\be \label{3char-mu1}
2\mu_1 =\frac49 - 8(\alpha_0 \alpha_1+ \alpha_1 \alpha_2+\alpha_0 \alpha_2), \hspace{1cm} 
\mu_2 = 8\alpha_0 \alpha_1 \alpha_2 
\ee 
Using Eqs.(\ref{3char-indicial}), (\ref{3char-mu1}) and (\ref{3char-n1}) and substituting for the Fourier coefficients of the Eisenstein series (see the Appendix) we get:
\be \label{m1 for 3char}
m^{(i)}_1 =\frac{24\alpha_i\big(20\alpha^2_i+(62\alpha_j -11)\alpha_i +62\alpha^2_j -31\alpha_j+1\big)}{(\alpha_i - \alpha_j +1)(4\alpha_i+2\alpha_j+1)},\quad j\neq i
\ee
Note that for any chosen $i$, this equation holds for both values of $j$ different from $i$. 

Let us now specialize to the case of identity character ($i=0$) and ask under what circumstances $m_1^{(0)}$ vanishes. By requiring \eref{m1 for 3char} to be zero we get the following relation between $\alpha_0$ and one of the other exponents, say $\alpha_1$:
\be \label{3char-alpha0}
\alpha_0 = \frac{1}{40} \left(11-62 \alpha_1 \pm \sqrt{41+1116 \alpha_1-1116 \alpha_1^2}\right)
\ee 
Since $\alpha_0$ and $\alpha_1$ are rational linear combinations of the central charge and conformal dimensions of one of the primaries (both of which are rational), we conclude that they themselves are rational numbers. It follows that the discriminant in \eref{3char-alpha0} is the square of some rational number $p$. Solving this equation for $\alpha_1$ we get:
\be
\alpha_1 = \frac{1}{186}\left(93 \pm \sqrt{31}\sqrt{320-p^2}\right)
\label{alphap}
\ee 
The rationality of $\alpha_1$ forces $\sqrt{320-p^2}$ to be of the form $\sqrt{31}\,q$, where $q$ is some rational number. Squaring both sides of this equation gives us the following Diophantine equation:
\be \label{3char-diophantine}
p^2+ 31q^2=320
\ee
This equation describes an ellipse. Thus, we see that rational points $(p,q)$ on this ellipse correspond to possible candidate $\alpha$'s describing 3-character theories with $\ell=0$ and without a Kac-Moody algebra. Of course these candidates, if found, would only have passed a low-level test and we would then have to determine their characters and check integrality of their coefficients to high orders before having any confidence that they exist as CFT's. This check is straightforward to perform because for $\ell=0$ and three characters, the exponents completely specify the differential equations and thereby the characters.

We already know one solution to the above requirements that is definitely a CFT, namely the Ising model. This has $c=\half$ and conformal dimensions $\frac12, \frac{1}{16}$. It is easy to verify that the characters of this theory have $\ell=0$, which as we already pointed out is the case for all minimal models. And it has $m_1^{(0)}=0$, because minimal models have no Kac-Moody symmetry. We will find it useful to start by describing the Ising model as a rational point of the ellipse of \eref{3char-diophantine}. Indeed using Eqs.(\ref{alphap}) and (\ref{3char-diophantine}) we easily find that it represents the point $(p_0,q_0)=(\frac{37}{4},\frac{11}{4})$ on this ellipse.

Using this as a ``base point'' we will search for other rational points on the ellipse. Let us consider a line with variable slope passing through the point $(p_0,q_0)$ and look for rational points where it intersects the ellipse. 
A line through $(p_0,q_0)$ can be parametrised as follows:
 \be \label{pq}
( p,q)= (p_0 -\gamma t,q_0-t)
 \ee
where $\gamma$ is a real parameter. Given that $(p_0,q_0)$ are rational, $(p,q)$ will also be rational if $t, \gamma t$ are rational. This means that $\gamma$ in particular must be rational. Now substituting the above in \eref{3char-diophantine} permits us to solve for $t$ in terms of $p_0,q_0$ and $\gamma$. Putting this back in the above, we get (after excluding $t=0$):
\be \label{pqexpr}
(p,q) = \Bigg( p_0-\gamma \left (\frac{2 \gamma p_0 + 62 q_0}{31 + \gamma^2}\right),q_0- \left (\frac{2 \gamma p_0  + 62 q_0}{31 + \gamma^2}\right) \Bigg)
\ee
Thus we have solved the initial problem, that of finding all rational points on the ellipse. There is one such point for every rational $\gamma$.

Next we use the recursive solution to find the second-level degeneracy in terms of the exponents $\alpha_i$:
\be
\begin{split}
	m^{(i)}_2&=\Bigg[36 \alpha_i \bigg(2 + 3200 \alpha_i^5 - 339 \alpha_j + 2897 \alpha_i^2 - 8876 \alpha_j^3 + 8876 \alpha_j^4 + 80 \alpha_i^4 (-1 + 248 \alpha_j)\\
	&\quad + 8 \alpha_i^3 (-277 - 718 \alpha_j + 6324 \alpha_j^2) + \alpha_i^2 (953 - 6136 \alpha_j- 17700 a1^2 + 61504 \alpha_j^3) + \\
   &\qquad	\alpha_i(-109 + 2702 \alpha_j - 5236 \alpha_j^2 - 13000 \alpha_j^3 + 
30752 \alpha_j^4)\bigg)\Bigg]\times\\
&\qquad \Big[(1 + \alpha_i - \alpha_j) (1 + 4 \alpha_i + 2 \alpha_j) (2+\alpha_i-\alpha_j)(3+4\alpha_i+2\alpha_j)\Big]^{-1}
\end{split} 
\label{3char-m2}
\ee
Using Eqs.(\ref{alphap}) and (\ref{3char-alpha0}) we can express the exponents $\alpha_i$  in terms of $p,q$:
\be \label{alphaexpr}
(\alpha_0,\alpha_1,\alpha_2) = \Big(-\half+\frac{31q-3p}{120},\half-\frac{q}{6},\frac12+\frac{(3p-11q)}{120} \Big)
\ee 
Substituting the values of this in \eref{3char-m2}, we obtain $m_2$ (as usual, this quantity without a superscript refers to the identity character) as a rational function of $\gamma$:
\be
m_2= \frac{-(-5363 - 62 \gamma + 53 \gamma^2) (18480991 + 359538 \gamma - 
116516 \gamma^2 - 1058 \gamma^3 + 
21 \gamma^4)}{4 (31 + \gamma^2) (-31 - 9 \gamma + 
6 \gamma^2) (-527 + 82 \gamma + 97 \gamma^2)}
\label{mtwoeq}
\ee

Our strategy is now to consider all rational points on the ellipse, i.e. all rational numbers $\gamma$, and ask which ones specify a non-negative integer $m_2$ via \eref{mtwoeq}. If they give fractional or negative $m_2$, they can be eliminated. This procedure will rule out all but a small number of cases. Accordingly, we searched for all rational solutions to \eref{mtwoeq} for values of $m_2$ ranging from 1 to 2000,000. Solutions are quite sparse, with only nine possible values of $m_2$ in the range 1 to 100,000 and not a single one after that. We suspect (but cannot rigorously prove) that these are all the solutions. 

We found six rational values of $\gamma$ for $m_2=1$, and two for each of the other allowed values of $m_2$. Once we have the values of $\gamma$ that solve \eref{mtwoeq}, we use equations \eref{pqexpr} and \eref{alphaexpr} to obtain the exponents $\alpha_i$. The central charges for these candidates can be computed as $c=-24\alpha_0$. 
It turns out that there are two different values of $\gamma$ for each set of exponents $\alpha_i$, with the roles of $\alpha_1$ and $\alpha_2$ exchanged. This can be explained by observing that  \eref{alphaexpr} has a symmetry under the transformation $(p,q) \rightarrow (-\frac{93q}{20}-\frac{11p}{20},\frac{11q}{20}-\frac{3p}{20})$ which leaves $\alpha_0$ unchanged but exchanges $\alpha_1$ with $\alpha_2$. Therefore, for $m_2=1$ the six different values of $\gamma$ group into three pairs corresponding to three different sets of exponents $\alpha_i$. There is a candidate 3-character theory for each set. On the other hand for $m_2\ge 2$ we have a single set of exponents for each pair of $\gamma$ values, and therefore a single candidate theory. The results at this stage are exhibited in Table \ref{table-m2}.

\begin{table}[ht]
	\centering
	\begin{tabular}{|c|c|c| |c|c|c|c|}
     	\hline
		No. & $m_2^{(0)}$ & $\gamma$ & $\alpha_0$ & $\alpha_1$ & $\alpha_2$ & $c=-24\alpha_0$ \\
		\hline
		1 &$1$ & $-13,15$  & $\phantom{-}\frac{11}{60}$ & $-\frac{1}{60}$ & $\frac{1}{3}$ & $-\frac{22}{5}$   \\
		\hline
		2  &  $1$ &  $-33, 47$  &$\phantom{-}\frac{17}{42}$& $-\frac{1}{42}$  & $\frac{5}{42}$ & $-\frac{68}{7}$  \\
				\hline
		3 & 1 & $-\frac{341}{37},\frac{31}{3}$  & $-\frac{1}{48}$ & $\phantom{-}\frac{1}{24}$ &  $\frac{23}{48}$& $\phantom{-}\frac12$  \\
		\hline
		4 & 2 & $-\frac{217}{9},31$  & $\phantom{-}\frac{11}{30}$ & $-\frac{1}{30}$ &  $\frac{1}{6}$& $-\frac{44}{5}$  \\
		\hline
		5 & 156 & $7,-\frac{19}{3}$  & $-\frac{1}{3}$ & $\phantom{-}\frac{2}{3}$ &  $\frac{1}{6}$& $\phantom{-} 8$  \\
		\hline
		6 & 2296 & $-\frac{31}{7},\frac{93}{19}$  & $-\frac{2}{3}$ & $\phantom{-}\frac{1}{3}$ &  $\frac{5}{6}$& $\phantom{-}16$  \\
		\hline
		7 & 49291 & $\frac{1}{3},-\frac{1}{17}$  & $-\frac{121}{84}$ & $\phantom{-}\frac{83}{84}$ &  $\frac{20}{21}$& $\phantom{-}\frac{242}{7}$  \\
		\hline
		8 & 63366 & $\frac{31}{33},-\frac{31}{47}$  & $-\frac{59}{42}$ & $\phantom{-}\frac{43}{42}$ &  $\frac{37}{42}$& $\phantom{-}\frac{236}{7}$  \\
		\hline
		9 & 63428 & $-\frac{403}{131},\frac{31}{9}$  & $-\frac{101}{105}$ & $\phantom{-}\frac{107}{210}$ &  $\frac{20}{21}$& $\phantom{-}\frac{808}{35}$  \\
		\hline
		10 & 90118 & $\frac97,-1$ &  $-\frac{41}{30}$ & $\phantom{-}\frac{31}{30}$ &  $\frac{5}{6}$& $\phantom{-}\frac{164}{5}$ \\
		\hline
		11 & 96256 & $\frac{37}{11},-3$  & $-\frac{47}{48}$ & $\phantom{-}\frac{23}{24}$ &  $\frac{25}{48}$& $\phantom{-}\frac{47}{2}$ \\
		\hline
		
	\end{tabular}
	\caption{Solutions to \eref{mtwoeq}. The exponents $\alpha_i$ are obtained using Eqs.(\ref{pqexpr}) and (\ref{alphaexpr}). The given values of $\alpha_1,\alpha_2$ correspond to the first value of $\gamma$ exhibited, while they are interchanged if we use the second value.}
	\label{table-m2}
\end{table}

We now try to understand whether these candidates really exist as characters, and if so, to what CFT's they are associated. First of all given the exponents $\alpha_i$ in Table \ref{table-m2}, we check the absence of any spin-1 primary (recall that the primary dimension is $h_i=\alpha_i-\alpha_0,~ i=1,2$). This rules out lines 5 and 6 of the table. Next using these exponents
we can evaluate the corresponding character as a $q$-series using the modular-invariant differential equation. Thereby we check to very high orders that $m_n^{(0)}$ are non-negative integers. We also verify that $m_n^{(i)}$ for the other two characters is a non-negative rational number. We reject candidates that do not satisfy these consistency conditions. Using these criteria and carrying out this analysis on each line of Table \ref{table-m2}, we find that the entries for $m_2^{(0)}=49291$ and 63428 must also be rejected. The surviving candidates are those appearing in lines $1-4, 8, 10, 11$ of the table.

Examining the exponents for these cases, we easily see that the cases in lines $1-4$ of the table correspond to known CFT's. The first three are, respectively, the minimal models for $(p,p')=(2,5),(2,7)$ and $(3,4)$ while the fourth one is the tensor product of two copies of the $(2,5)$ minimal model. Notice that case 1 is really a 2-character theory that has appeared as the solution of a 3-character differential equation (this means it has a spurious ``third character'' with which the first two do not mix under modular transformations, much as for the $E_8$ case discussed in Ref.\cite{Mathur:1988na}). This was already discussed as a 2-character theory in the previous section and we therefore ignore it in the present discussion. The others are all genuine 3-character theories. One of them, with $c=\half$, is the famous Ising model. Since these theories exist and satisfy all the criteria for which we have been searching, it is of course reassuring to find them. But the important question is whether there are any more. Remarkably it turns out that there are {\em precisely three} more theories in our list that are {\em not} minimal models. Moreover, they precisely satisfy the bilinear ``dual'' relation to the 3-character minimal models given in \eref{jbilinear} with $p=3$.

To see the relations between the new and old cases, let us compare lines 2 and 8 of Table \ref{table-m2}. Note first that the central charges add up to 24. Next, each of the conformal dimensions $h_1=\alpha_1-\alpha_0$ and $h_2=\alpha_2-\alpha_0$ for these two lines adds up to 2. These are precisely the properties of the bilinear relation (and also of the novel coset construction of Ref.\cite{Gaberdiel:2016zke} to which it is analogous). The same properties hold when we compare lines 3 and 11 of the table, and lines 4 and 10. Thus we have found three more pairs of theories that may potentially satisfy a bilinear relation giving the Moonshine CFT.

Again the proposed relations can be verified to high orders in $q$. We simply compute the characters of the potentially related pairs in Table \ref{table-m2} using the differential equations approach, multiply them pairwise, add them up and compare with $J(q)$ to each order. As before, this verification involves an ambiguity in the normalisations of the non-identity characters, since these are not determined by the differential equation. Compared with the 2-character case in the previous section, here we are studying 3-character theories so there are two undetermined primary degeneracies $\tD_1,\tD_2$. These are determined by imposing the bilinear identity to the first two nontrivial orders in $q$. Thereafter we check whether the identity continues to hold up to $\calo(q^{1000})$. We find that each pair passes this test perfectly. One caveat is that we again encounter negative degeneracies when the original theory is non-unitary, so in these cases the duals are ``exotic'' so it may or may not be possible to make sense of them as some kind of CFT's. For the dual to $\calm_{2,7}$, denoted $\tilde\calm_{2,7}$, we have the exponents:
\be
(\talpha_0,\talpha_1,\talpha_2)=\frac{1}{42}(-59,43,37)
\ee
and the degeneracies of the non-identity primaries are $(\tD_1,\tD_2)=(-715139,848656)$. For the dual to $\calm_{2,5}\times \calm_{2,5}$ the exponents are:
\be
(\talpha_0,\talpha_1,\talpha_2)=\frac{1}{30}(-41,31,25)
\ee
and the degeneracies are $(\tD_1,\tD_2)=(615164,-508400)$. Notice that unlike the two-character case in the previous section, here {\em all} the degeneracies of the associated character are negative (equivalently, all are positive after extracting the overall negative degeneracy of the ground state). Thus it may be possible to make sense of these as theories with a fermion number\footnote{We thank Matthias Gaberdiel for this suggestion.}.

This time we also have a unitary case, the Ising model, for which everything works perfectly. The ''dual'' theory with which it obeys a bilinear identity has $c=\frac{47}{2}$ and the degeneracies all turn out to be non-negative integers. Again we have determined the degeneracies and verified that the bilinear identity holds to high orders in $q$. For this case the exponents are:
\be
(\talpha_0,\talpha_1,\talpha_2)=\frac{1}{48}(-47,46,25)
\ee
and the degeneracies are $(\tD_1,\tD_2)=(96256,4371)$.

All in all, there is now a strong case that every minimal model with two or three characters has an associated ``dual'' CFT (exotic when the original theory is non-unitary, but normal when the original is unitary) which pairs with it to give the Moonshine Module. The resulting pairings are summarised in Table \ref{table-3char}. It is amusing to note that the values of $\gamma$ for each pair of models satisfying the bilinear relation are related by the inversion $\gamma\to-\frac{31}{\gamma}$.

\begin{table}[ht]
	\centering
	\begin{tabular}{|c|c|c|c|c|c|}
     	\hline
		No & $m_2^0$ & $\gamma$ & $D_1$ &$D_2$ & Identification\\
		\hline\hline
		1  &  $1$ &  $-33,47$ & $1$ & $1$ & $\cM_{2,7}$ 
		minimal model \\
			\hline
		2 & 63366 & $\frac{31}{33},-\frac{31}{47}$  & $-715139$ &  $848656$ & Dual of $\cM_{2,7}$ \\		
				\hline\hline
		3 & 1 & $-\frac{341}{37},\frac{31}{3}$ & $1$ & $1$ & $\calm_{3,4}$ (Ising model) \\
		\hline
		4 & 96256 & $\frac{37}{11},-3$ & $96256$ & $4371$ &  Dual of $\calm_{3,4}$ \\
		\hline\hline
		5 & 2 & $-\frac{217}{9},31$ & $1$ & $1$ & $\cM_{2,5}\otimes \cM_{2,5}$  \\
				\hline
		6 & 90118 & $\frac97,-1,$ & $615164$ &  $-508400$ & Dual of $\cM_{2,5}\otimes \cM_{2,5}$ \\
				\hline
		
	\end{tabular}
	\caption{$\ell=0$ three-character theories without Kac-Moody algebra. Here $m_2^{(0)}$ is the degeneracy of the second excited state in the identity character, $\gamma$ is the rational number in \eref{pq} and 
	$D_1$, $D_2$ are the ground-state degeneracies of the non-trivial primaries.}
	\label{table-3char}
\end{table}

Again it is worth pointing out that, independent of the bilinear relation, we have successfully classified all possible three-character RCFT's with $\ell=0$ having no Kac-Moody algebra. There are precisely three such theories, one of them a normal CFT (which we discuss in the following section) and the other two ``exotic'' in the sense of having negative, but integer, degeneracies. 

\subsection{Relation to the Baby Monster}

Clearly it is important to understand the ``new'' theories, namely entries $2,4,6$ of Table \ref{table-3char}. We note that these candidates have large values of $m_2$. They also have large degeneracies for the non-identity primary. Since two of them are exotic (and therefore may or may not exist as CFT's) we will focus on the sole unitary candidate, the one related to the Ising model. One of its fascinating features is that the number 96256 occurs twice: once as the degeneracy of the second excited state in the identity character ($m_2$) and once as the degeneracy of one of the nontrivial primaries. The number 4371 is the dimension of the other primary. Both these numbers are related to the Baby Monster, the second largest sporadic group. 

Indeed, there is a Baby Monster Vertex Operator Algebra $V\IB^\natural$ \cite{Hoehn:thesis} with central charge $\frac{47}{2}$ whose character (the ``shorter Moonshine module'') has the expansion:
\be
\chi_{V\IB^\natural}=q^{-\frac{47}{48}}\left(1+4371 q^{\frac32} +96256 q^2 
+1143745 q^{\frac52}+\cdots\right)
\ee
This can be rewritten:
\be
\begin{split}
\chi_{V\IB^\natural}&=q^{-\frac{47}{48}}\left(1 +96256 q^2+\cdots\right)+
 q^{\frac{25}{48}}\left(4371+ 1143745 q+\cdots\right)\\
 &=\chi_0(q)+\chi_2(q)
\end{split}
\ee
where the second line is a sum of two of the characters of our 3-character $c=\frac{47}{2}$ theory. Moreover the three characters $\tchi_0,\tchi_1,\tchi_2$ of the $c=\frac{47}{2}$ theory discussed above appear in Eq.(4.13) of Ref.\cite{Hoehn:thesis} (see also Ref.\cite{Hoehn:Baby8}) as the modules $V\IB^\natural(0),V\IB^\natural(2),V\IB^\natural(1)$ respectively. The bilinear relation \eref{jbilinear} for this case will then decompose the dimensions of representations of the Monster in terms of those of the Baby Monster. 

The dual relation between the Ising model and the above theory seems  to originate in the fact that the Moonshine Module is itself a sum of irreducible characters of 48 copies of the Ising model, as shown in Ref.\cite{Dong:1994}\footnote{We thank Matthias Gaberdiel for bringing this fact and this reference to our attention.}. Indeed this observation was relevant for the subsequent work of H\"ohn on the $c=\frac{47}{2}$ ``Baby Monster'' module.
Nevertheless, the way we have discovered the Baby Monster as a three-character RCFT without a Kac-Moody algebra, which pairs with Ising model to give the Moonshine character, could provide a simple method to reproduce some of the results of Ref.\cite{Hoehn:thesis}, add new perspectives and give rise to generalisations.

Our results lead one to ask whether other minimal models have duals of the sort we have found. The next minimal model, $\calm_{4,5}$ (the tri-critical Ising model) has 6 characters and a central charge of $\frac{7}{10}$. If it has a dual in our sense (with which it obeys a bilinear identity giving the Moonshine Module) then that theory must have six characters as well, and a central charge $c=\frac{233}{10}$. It is easy to verify using $\alpha_0+\talpha_0=-1,\alpha_i+\talpha_i=1, i=1,\ldots,5$ that the dual has $\ell=6$, i.e. the coefficients in its differential equation can have $\frac{\ell}{6}=1$ full singularity in moduli space. This allows a rather large number of independent coefficient functions, hence the free parameters in the differential equation cannot be determined completely by the conjectured exponents $\talpha_i$. Thus we have no obvious way of generating the $q$-series for the possible dual characters and verifying their integrality. The situation rapidly gets even more complicated for other unitary minimal models. In short, for the moment we have no way to support nor exclude the existence of such duals for minimal models. 

\section{Summary and Discussion}

In this section we summarise the emerging picture and propose possible lines of further investigation. We have classifed all two-character RCFT's with no Kac-Moody algebra for $\ell\le 2$ ($\ell=0,3$ were already done) and all three-character RCFT's with the same property for $\ell=0$ (the latter, under the assumption that our computation upto $m_2= 2,000,000$ is sufficient). The restriction to low values of $\ell$ is due to the fact that the method of modular-invariant differential equations is most restrictive in these cases and allows efficient construction of candidate characters given only the critical exponents. 

Within this set of systems, we found the expected unitary and non-unitary minimal models (precisely those with two and three characters) as well as dual theories in every case, although the duals for non-unitary minimal models have negative integer degeneracies. Each model and its dual satisfies a bilinear pairing identity equating it to the Moonshine Module. This pairing is reminiscent of that recently found in Ref.\cite{Gaberdiel:2016zke} between an affine theory and the coset of a meromorphic theory by that affine theory, although affine Lie algebras play no role in the present case -- by construction. It would be interesting to know if more such pairs of theories exist. This structure is interesting from the mathematical point of view as well, since the unitary case we discovered in our approach is a known theory related to the Baby Monster module, and our pairing decomposes representations of the Monster into the Baby Monster.

\section*{Acknowledgements}

We would like to thank R. Loganayagam and Sameer Kulkarni for useful discussions and Matthias Gaberdiel for very helpful comments on a first draft of this manuscript. The work of HRH is supported by an INSPIRE Scholarship, DST, Government of India, and that of SM by a J.C. Bose Fellowship, DST, Government of India. SM would also like to acknowledge the warm hospitality of the Yukawa Institute of Theoretical Physics, Kyoto University where this work was completed. 
We thank the people of India for their generous support for the basic sciences.

\section*{Conventions and useful formulae}

The relevant Eisenstein series used in this paper, normalised so that their first term is unity, have the series expansion:
\be
\begin{split}
E_2 &= 1 -24\sum_{n=1}^\infty \frac{n q^n}{1-q^n}=1-24\sum_{n=1}^\infty \sigma_1(n)q^n\\
E_4 &= 1 +240\sum_{n=1}^\infty \frac{n^3 q^n}{1-q^n}=1+240\sum_{n=1}^\infty \sigma_3(n)q^n\\
E_6 &= 1 -504\sum_{n=1}^\infty \frac{n^5 q^n}{1-q^n}=1-504\sum_{n=1}^\infty \sigma_5(n)q^n\nn
\end{split}
\ee
where
\be
\sigma_p(n)=\sum_{d|n}d^p\nn
\ee
$E_4$ and $E_6$ can be expressed in terms of Jacobi $\theta$-functions:
\be
\begin{split}
E_4&=\half \sum_{\nu=2}^4 \big(\theta_\nu(0|\tau)\big)^8\\
E_6&= \sqrt{E_4^3-\sfrac{27}{4}(\theta_2\theta_3\theta_4)^8}\nn
\end{split}
\ee
The explicit expansion of these series  to a few finite orders is:
\be
\begin{split}
E_2 &= 1-24q -72 q^2 - 96 q^3 - 168 q^4\\
E_4 &= 1+240 q+2160 q^2+6720 q^3+17520 q^4\\
E_6 &= 1- 504q-16632 q^2 - 122976 q^3-532728 q^4\nn
\end{split}
\ee


\bibliographystyle{JHEP}
\bibliography{m1eq0}

\providecommand{\href}[2]{#2}\begingroup\raggedright\begin{thebibliography}{10}

\bibitem{DiFrancesco:1997nk}
P.~Di~Francesco, P.~Mathieu, and D.~Senechal, {\em {Conformal Field Theory}}.
\newblock Graduate Texts in Contemporary Physics. Springer-Verlag, New York,
  1997.

\bibitem{Brown:1986nw}
J.~D. Brown and M.~Henneaux, {\it {Central Charges in the Canonical Realization
  of Asymptotic Symmetries: An Example from Three-Dimensional Gravity}},  {\em
  Commun. Math. Phys.} {\bf 104} (1986) 207--226.

\bibitem{Goddard:1984vk}
P.~Goddard, A.~Kent, and D.~I. Olive, {\it {Virasoro Algebras and Coset Space
  Models}},  {\em Phys. Lett.} {\bf B152} (1985) 88--92.

\bibitem{Goddard:1986ee}
P.~Goddard, A.~Kent, and D.~I. Olive, {\it {Unitary Representations of the
  Virasoro and Supervirasoro Algebras}},  {\em Commun. Math. Phys.} {\bf 103}
  (1986) 105--119.

\bibitem{Belavin:1984vu}
A.~A. Belavin, A.~M. Polyakov, and A.~B. Zamolodchikov, {\it {Infinite
  Conformal Symmetry in Two-Dimensional Quantum Field Theory}},  {\em Nucl.
  Phys.} {\bf B241} (1984) 333--380.

\bibitem{Andrews:1984af}
G.~E. Andrews, R.~J. Baxter, and P.~J. Forrester, {\it {Eight vertex SOS model
  and generalized Rogers-Ramanujan type identities}},  {\em J. Statist. Phys.}
  {\bf 35} (1984) 193--266.

\bibitem{Frenkel:1988xz}
I.~Frenkel, J.~Lepowsky, and A.~Meurman, {\em Vertex Operator Algebras and the
  Monster}.
\newblock Academic Press, Boston, USA, 1988.

\bibitem{Anderson:1987ge}
G.~Anderson and G.~W. Moore, {\it {Rationality in Conformal Field Theory}},
  {\em Commun. Math. Phys.} {\bf 117} (1988) 441.

\bibitem{Eguchi:1988wh}
T.~Eguchi and H.~Ooguri, {\it {Differential Equations for Characters of
  Virasoro and Affine Lie Algebras}},  {\em Nucl. Phys.} {\bf B313} (1989) 492.

\bibitem{Mathur:1988rx}
S.~D. Mathur, S.~Mukhi, and A.~Sen, {\it {Differential Equations for
  Correlators and Characters in Arbitrary Rational Conformal Field Theories}},
  {\em Nucl. Phys.} {\bf B312} (1989) 15.

\bibitem{Mathur:1988na}
S.~D. Mathur, S.~Mukhi, and A.~Sen, {\it {On the Classification of Rational
  Conformal Field Theories}},  {\em Phys. Lett.} {\bf B213} (1988) 303.

\bibitem{Mathur:1988gt}
S.~D. Mathur, S.~Mukhi, and A.~Sen, {\it {Reconstruction of Conformal Field
  Theories From Modular Geometry on the Torus}},  {\em Nucl. Phys.} {\bf B318}
  (1989) 483.

\bibitem{Hampapura:2015cea}
H.~R. Hampapura and S.~Mukhi, {\it {On 2d Conformal Field Theories with Two
  Characters}},  {\em JHEP} {\bf 01} (2016) 005,
  [\href{http://xxx.lanl.gov/abs/1510.0447}{{\tt arXiv:1510.0447}}].

\bibitem{Gaberdiel:2016zke}
M.~R. Gaberdiel, H.~R. Hampapura, and S.~Mukhi, {\it {Cosets of Meromorphic
  CFTs and Modular Differential Equations}},  {\em JHEP} {\bf 04} (2016) 156,
  [\href{http://xxx.lanl.gov/abs/1602.0102}{{\tt arXiv:1602.0102}}].

\bibitem{Hoehn:thesis}
G.~{Hoehn}, {\it {Selbstduale Vertexoperatorsuperalgebren und das Babymonster
  (Self-dual Vertex Operator Super Algebras and the Baby Monster)}},  {\em
  ArXiv e-prints} (June, 2007) [\href{http://xxx.lanl.gov/abs/0706.0236}{{\tt
  arXiv:0706.0236}}].

\bibitem{Naculich:1988xv}
S.~G. Naculich, {\it {Differential Equations for Rational Conformal
  Characters}},  {\em Nucl. Phys.} {\bf B323} (1989) 423.

\bibitem{Schellekens:1992db}
A.~N. Schellekens, {\it {Meromorphic C = 24 conformal field theories}},  {\em
  Commun. Math. Phys.} {\bf 153} (1993) 159--186,
  [\href{http://xxx.lanl.gov/abs/hep-th/9205072}{{\tt hep-th/9205072}}].

\bibitem{Hoehn:Baby8}
G.~{Hoehn}, ``{Generalized Moonshine for the Baby Monster}.''
  {https://www.math.ksu.edu/$\sim$gerald/papers/baby8.ps}.

\bibitem{Dong:1994}
C.~Dong, G.~Mason, and Y.~Zhu, {\it {Discrete Series of the Virasoro Algebra
  and the Moonshine Module}},  in {\em Proceedings of Symposia in Pure
  Mathematics Vol 56, Part 2: Algebraic Groups and Their Generalizations:
  Quantum and Infinite-Dimensional Methods} (W.~J. Haboush and B.~J. Parshall,
  eds.), pp.~295--316.
\newblock American Mathematical Society, 1994.

\end{thebibliography}\endgroup

\end{document}